# High-efficiency single photon emission from a silicon T-center in a nanobeam

Chang-Min Lee,[1,2] Fariba Islam,[1,2] Samuel Harper,[1,2] Mustafa Atabey Buyukkaya,[1,2] Daniel Higginbottom,[3,4] Stephanie Simmons,[3,4] and Edo Waks[1,2,*]

[1]Institute for Research in Electronics and Applied Physics and Joint Quantum Institute, University of Maryland, College Park, Maryland 20742, USA
[2]Department of Electrical and Computer Engineering, University of Maryland, College Park, Maryland 20740, USA
[3]Department of Physics, Simon Fraser University, Burnaby, BC V5A 1S6, Canada
[4]Photonic Inc., Coquitlam, BC, Canada

**Abstract:**
Color centers in Si could serve as both efficient quantum emitters and quantum memories with long coherence times in an all-silicon platform. Of the various known color centers, the T center holds particular promise because it possesses a spin ground state that has long coherence times. But this color center exhibits a long excited state lifetime which results in a low photon emission rate, requiring methods to extract photon emission with high efficiency. We demonstrate high-efficiency single photon emission from a single T center using a nanobeam. The nanobeam efficiently radiates light in a mode that is well-matched to a lensed fiber, enabling us to collect over 70% of the T center emission directly into a single mode fiber. This efficiency enables us to directly demonstrate single photon emission from the zero phonon line, which represents the coherent emission from the T center. Our results represent an important step towards silicon-integrated spin-photon interfaces for quantum computing and quantum networks.

## Introduction

Optically active solid-state qubits are essential components for scalable quantum devices. These qubits serve as fundamental building blocks for distributed and modular quantum computers[1,2], quantum networks[3], and quantum sensors[4–6]. They can also be integrated with nanophotonic devices to enhance photon emission and attain strong light-matter interactions[2]. Significant progress has already been made in nanophotonic integration of multiple solid-state qubit platforms including color centers in diamond and SiC, quantum dots, and rare-earth ions[7].

Silicon is a compelling material system for solid-state qubit integration[1,5,7]. Silicon photonic devices achieve unparalleled scalability due to availability of large foundries capable of producing highly complex devices. Additionally, the most abundant silicon isotope, $^{28}$Si, is free of nuclear spin, thus creating a favorable spin environment that can be further improved by isotopic purification[8–10]. However, engineering optically active spin qubits in silicon faces significant challenges because silicon is a narrowband indirect semiconductor that typically exhibits poor optical efficiency.

Color centers in silicon have recently emerged as a potential solution to the problem. Single photon emission from the G center has been demonstrated by a number of groups[11–17]. But the G center does not possess an unpaired electron and therefore does not act as a stable spin qubit. In contrast, the T center does possess a spin ground state and is therefore a promising optically active spin qubit[18–22] with millisecond coherence times[18]. Moreover, it emits photons in the telecom O-band, making it suitable for silicon photonic integration and long-distance propagation through optical fibers. The properties of T centers have already been investigated in SOI (silicon-on-insulator), which offers the appropriate layer structure for nanophotonic integration[19–21]. However, the T center has a relatively long excited state lifetime compared to the G center, making it a dim emitter. Consequently, most research efforts have focused on collecting emission from the ensemble of the T center[18,19,21,22], or from the phonon sideband of a single T center which is significantly brighter than the zero phonon line[20]. But the zero phonon emission is particularly significant because it represents the coherent emission, which is essential for spin-photon entanglement and distribution of quantum information over longer distances[23]. To make use of this quantum emission requires efficient collection methods.

In this work we demonstrate efficient collection from the zero phonon line of a single T center in a silicon nanobeam. We achieve a coupling efficiency from the nanobeam into a single mode fiber exceeding 70%. This enhanced efficiency enables us to collect emission directly from the zero phonon line and validate its single photon nature. We measure a photon count rate from the zero phonon line exceeding 1 kHz, brighter than previously reported from a single T center. Our work represents an important step towards efficient integration of silicon spin qubits with silicon nanophotonics for scalable quantum information processing.

**Physical system and fabricated device**

We implanted T centers in a four inch silicon-on-insulator (SOI) wafer by a series of implantation and annealing steps[19]. We utilized an initial commercial SOI wafer (WaferPro) with a device layer thickness of 220 ± 10 nm and a (100) crystal orientation. To penetrate approximately 110 nm of Si, we employed C-12 ion implantation at an energy of 38 KeV and the fluence of $7\times10^{12}$ ions/cm$^2$ (CuttingEdge Ions). After the first implantation, the wafers underwent rapid thermal annealing at 1000°C for 20 seconds in an Ar environment. To achieve the same depth, hydrogen was then implanted at 9 KeV, with the fluence of H ions being the same as that of C-12. Following the second implantation, the sample was immersed in deionized water and boiled for 1 hour. Subsequently, a second round of rapid thermal annealing was carried out at 400°C for 3 minutes in a nitrogen environment.

To efficiently collect the emission from a single T center, we utilized the tapered nanobeam structure[24] illustrated in Figure 1(a). The structure is composed of a nanobeam with width $b$, followed by a tapered region of length $L_{taper}$ that linearly reduces the nanobeam width down to $b_{taper}$. This tapered region expands the propagating mode so that it matches the transverse mode of lensed fiber that collects the emission from the edge. This expansion is shown in figure 1(b) which plots the calculated field along different segments along the nanobeam taper. A periodic array of air holes with radius $r$ and periodicity $a$ at the opposite end of the nanobeam act as a photonic crystal mirror, ensuring all emission is directed towards the tapered end.

We designed the taper to optimally couple to a lensed fiber with a numerical aperture of 0.4. From finite-difference time-domain simulations, we determined the optimal nanobeam parameters to be $a = 361$ nm, $r = 0.3a$, $b = 420$ nm, $b_{tip} = 150$ nm, and $L_{taper} = 12$ µm. The resulting transverse mode is shown in Figure 1(c). From finite-difference time-domain simulations we determined that for a dipole located inside the nanobeam and oriented in the transverse electric (TE) direction, 96% of the emission couples to the nanobeam waveguide mode and 90% of the emission is collected into the lensed fiber. We note that this number assumes perfect alignment of the dipole with the TE mode, and that in real devices this efficiency could be degraded due to imperfect spatial alignment and angular orientation.

To fabricate the nanobeam structure, we used a combination of electron beam lithography and transfer print lithography. We first utilized electron beam lithography and wet etch to pattern the SOI wafer, resulting in the tethered structure shown in Figure 2(a). To transfer the design onto the edge of the silicon wafer, we used a transfer print lithography method previously reported for transfer of InP nanobeams[24,25]. We utilized a polydimethylsiloxane (PDMS) stamp to pick a desired nanobeam from the patterned substrate, and place it on the edge of a separate silicon carrier wafer. Figure 2(b) shows an image of the final suspended structure. The beam is suspended from the edge of the carrier wafer, making it possible to collect the light from the edge using a lensed fiber.

**Results and discussion**

Figure 3(a) illustrates the experimental setup used for the measurements. We loaded both the silicon nanobeams containing T centers and a lensed fiber in a cryostat operating at 3.6 Kelvin. The nanobeams were fixed on a three-axis nanopositioner stage enabling them to move with respect to the lensed fiber. Both the nanobeam and the lensed fiber were installed on two nanopositioners so that the entire system moves with respect to the objective lens. To measure reflectivity of the nanobeam, we injected laser light through one end of a 90:10 fiber coupler and measured the power of the reflected light using a power meter (dashed box in Figure 3(a)). To excite the T center we injected an excitation laser at a wavelength of 780 nm from the top. The generated T center emission was collected using the lensed fiber. We measured the emission spectrum using a spectrometer and charge coupled device array. For photon counting and second-order autocorrelation, we utilized a fiber-integrated tunable filter with a bandwidth of 0.2 nm. The filtered signal was measured using the Hanbury-Brown and Twiss setup consisting of a 50:50 fiber coupler, two superconducting nanowire single photon detectors, and a time-correlated single photon counter.

We first performed reflectivity measurements to characterize the coupling efficiency of the nanobeam. We injected light into the nanobeam using a tunable external cavity diode laser emitting at 1326 nm wavelength and measured the light collected from the reflected port. By comparing the measured power at the input port and the reflected port, we measured the total round-trip reflectivity R to be 42%. Assuming the coupling-in and coupling-out efficiency are the same and taking into account the 90:10 fiber coupler transmission $T_{fc}$ of 83%, we determined the lensed fiber coupling efficiency to be η=71% ($R = \eta^2 T_{fc}$). We attribute the discrepancy of this efficiency from the simulated value to scattering loss at the tapered part of the nanobeam and the photonic crystal mirror, as well as potential angular mismatch between the nanobeam and lensed fiber.

Next, we measured the fluorescence of the sample using a continuous-wave above-band laser. We first excited all of the emitters in the nanobeam by shining a defocused above-band excitation laser at a wavelength of 780 nm on the entire nanobeam with a relatively large beam diameter of approximately 10 μm. We used an excitation power of 22 μW. Figure 3(b) shows the resulting spectrum. We observed two strong emission lines, one at 975 meV which corresponds to the zero phonon line emission from G centers and a second at 935 meV corresponding to zero phonon line emission from T centers (labeled as G and T in Figure 3(b), respectively). The G center peak is stronger than the T center peak due to the fact that the G center has two orders of magnitude shorter excited state lifetime[11,14,15], and also because of imperfect implantation which may generate a high concentration of G centers. Our measurements focused exclusively on the T center line.

In order to isolate a single T center, we pumped the nanobeam with a tightly focused laser from the top. We used the same pump laser but lowered the pump power to 1 μW due to tight focusing. Figure 3(c) shows the photoluminescence collected from the lensed fiber. We observed a single sharp peak at 935.4 meV corresponding to zero phonon line emission consistent with T centers[18–20]. A Lorentzian fit of the peak shows a linewidth of 41 μeV, which is close to the spectrometer resolution limit of 30 μeV (measured separately).

A single quantum emitter features saturation of the photon count rate as a function of pump power. To study the saturation response of the T center, we measured the count rates from the zero phonon line as a function of pump power (Figure 4(a)). We used a tunable fiber spectral filter with a bandwidth of 0.2 nm to isolate the emission of the zero phonon line. The filtered emission was recorded with superconducting nanowire single photon detectors. The measured data fit well to the characteristic two-level emitter saturation model $I(P) = I_{sat}P/(P + P_{sat}) + \alpha P$, where $I_{sat}$ and $P_{sat}$ are count rates at saturation and saturation pump power, respectively, and α is a coefficient for a weak linear background[15]. From the fit, we measured $I_{sat}$ = 2423 counts per second (cps) and $P_{sat}$ = 0.93 μW.

To further investigate the T center emission characteristics, we measured the second-order autocorrelation at different pump powers using the Hanbury Brown and Twiss setup (Figure 4(b)). Each histogram shows a clear dip at zero time delay, demonstrating that the emission is originating from a single photon source. To exclude the false counts generated by the background signal, we calculated background-corrected $g^{(2)}(\tau)$ by using the equation

$$g^{(2)}(\tau) = (C(\tau) - (1 - \rho^2))/\rho^2, \qquad (1)$$

where $C(\tau)$ is the normalized coincidence, $\rho = S/(S + B)$ is the signal to total counts ratio, S is the signal count rate, and B is the background count rate[26]. We measured the background count rate B by shifting the tunable filter 3 nm away from the zero phonon line. The signal to total counts ratio $\rho$ was measured to be from 0.76 to 0.87, as we increased the pump power from 0.3 μW to 1.5 μW. We fit the calculated $g^{(2)}(\tau)$ to a two-level emitter second order autocorrelation of $g^{(2)}(\tau) = 1 - (1 - b)e^{-\gamma_c \tau}$, where b is second order autocorrelation at zero time delay ($g^{(2)}(0)$), and $\gamma_c$ is the decay rate constant of the anti-bunching dip decay[27]. At the lowest pump power of 0.3 μW, we achieved $g^{(2)}(0) = 0.17(4)$.

The decay rate constant $\gamma_c$ depends on the excited state lifetime $\tau_{rad}$ and pump power P as $\gamma_c = 1/\tau_{rad}(1 + \beta P)$, where the second term corresponds to pump rate with the $\beta$ as a fitting parameter[27]. As we increased the pump power, the decay rate constant became faster because the pump rate increases (Figure 4(c)). By fitting the data to a linear function, we obtained the excited state lifetime $\tau_{rad} = 1.61\ \mu s$. This lifetime is in close agreement with previously reported T center lifetimes[18,20] and estimated lifetimes from first-principles calculation[28].

To estimate brightness of the T center, we analyzed the detected count rate along with the measured $g^{(2)}(0)$ values. In addition to the single photons emitted from the T center, the detected count rate includes other signals such as detector dark counts, the sideband of the G centers in the nanobeam, and photons emitted from other T centers in the nanobeam. Thus, the raw photon count rate represents an overestimation of the brightness of the T center. The correct single photon emission rate can be calculated using the equation[29]

$$I_{corr} = (I_{det} - B)\sqrt{1 - g^{(2)}(0)}, \tag{2}$$

where $I_{det}$ is the detected intensity from the T center and $I_{corr}$ is the corrected intensity which contains only the single photon emission from the T center. Figure 4(d) plots the corrected single photon count rate and $g^{(2)}(0)$ values at different pump powers. As we increased the pump power, the $g^{(2)}(0)$ values increased and $I_{corr}$ showed saturation behavior. At pump power of 1.5 μW, which is approximately 1.5×$P_{sat}$, the corrected count rate was measured to be 1090 cps.

We compare our measured count rate to those reported in previous work. Our measured and $g^{(2)}(0)$-corrected count rates of 1090 cps are 3.4 times larger than the previously reported values of 325 cps obtained from the phonon sideband[20]. It thus represents a substantial improvement in count rate of over an order of magnitude because the zero phonon line is also 3.4 times dimmer than the phonon sideband[18]. Combining these two factors we estimate over an order of magnitude improvement in overall photon collection efficiency.

**Conclusions**

In conclusion, we demonstrated bright single-photon emission from a silicon T center in a tapered nanobeam. The nanobeam achieved a collection efficiency of over 70%, resulting in the highest count rates reported from a T center to-date. The high efficiency of the device opens up the possibility for efficient optical access of the spin ground state, as demonstrated with other optically active spin qubits[6,30,31]. An important next step is to couple the zero phonon line to a cavity mode to enhance brightness and improve the Debye-Waller factor. Ultimately, our results are an important step towards scalable nanophotonic integration of Si color centers with nanophotonics for quantum networking and distributed quantum computing.

**Acknowledgments**

The authors acknowledge Kartik Srinivasan and late Gregory Simelgor at National Institute of Standard and Technology for providing capability of the rapid thermal annealer.
The Waks group would like to acknowledge financial support from the National Science Foundation(grants #OMA1936314, #OMA2120757, #PHYS1915375, and #ECCS1933546),

AFOSR grant #FA23862014072, the U.S. Department of Defense contract #H98230-19-D-003/008, and the Maryland-ARL QuantumPartnership (W911NF1920181).

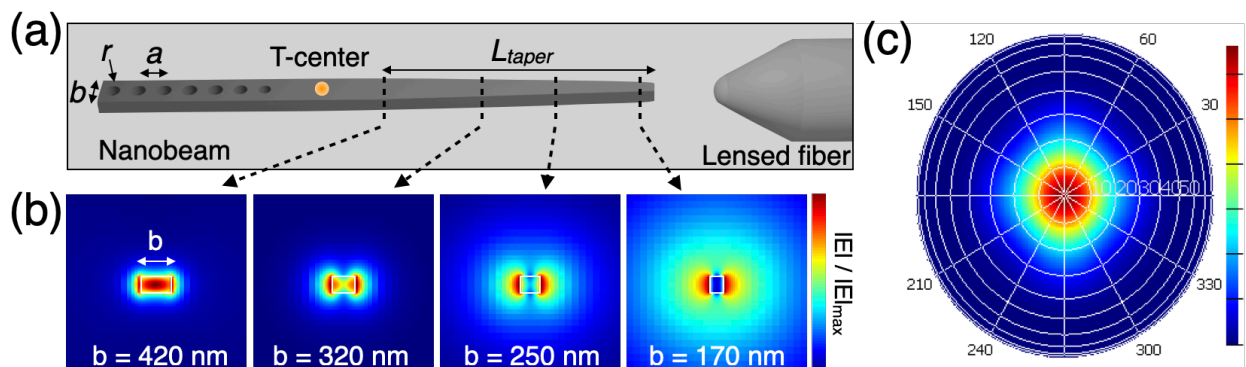

Figure 1 (a) A schematic of a silicon nanobeam containing a T center coupled to a lensed fiber. The dimensions of the nanobeam and the lensed fiber are not to scale. (b) Electric field magnitude profile at different cross sections (marked in panel (a)) of the nanobeam taper. White rectangles indicate the size of the nanobeam cross sections. (c) Far-field profile of the tapered nanobeam.

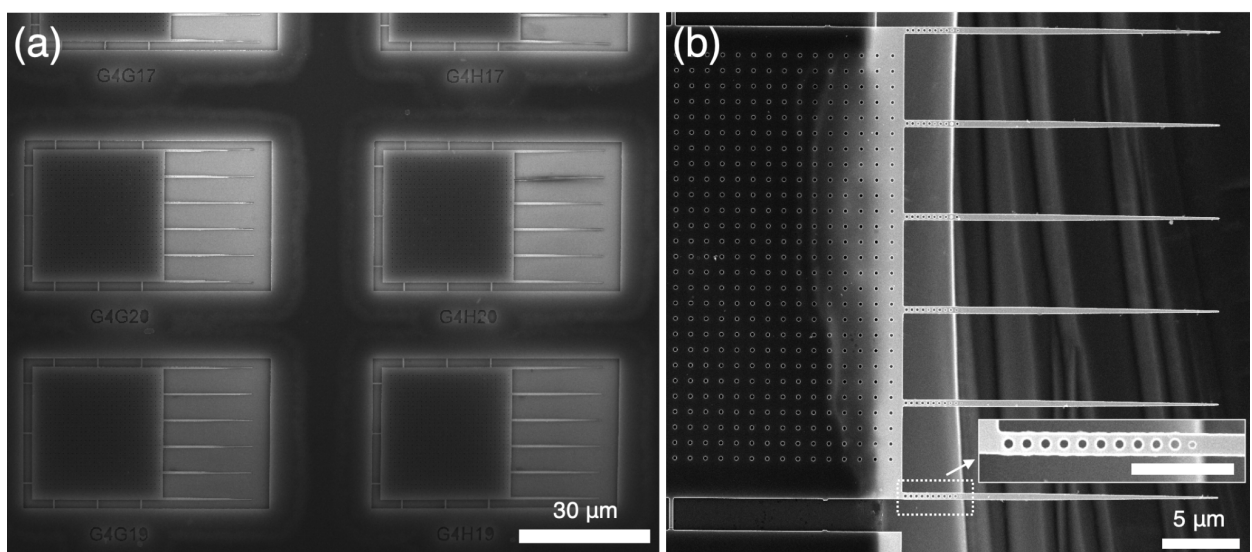

Figure 2 (a) A scanning electron micrograph of nanobeam arrays after e-beam lithography and wet etching. (b) A scanning electron micrograph of the nanobeam array that is transferred to a silicon carrier chip. Inset is a zoomed-in image of a photonic crystal mirror portion of the nanobeam (white dashed box). Scale bar of the inset is 2 μm.

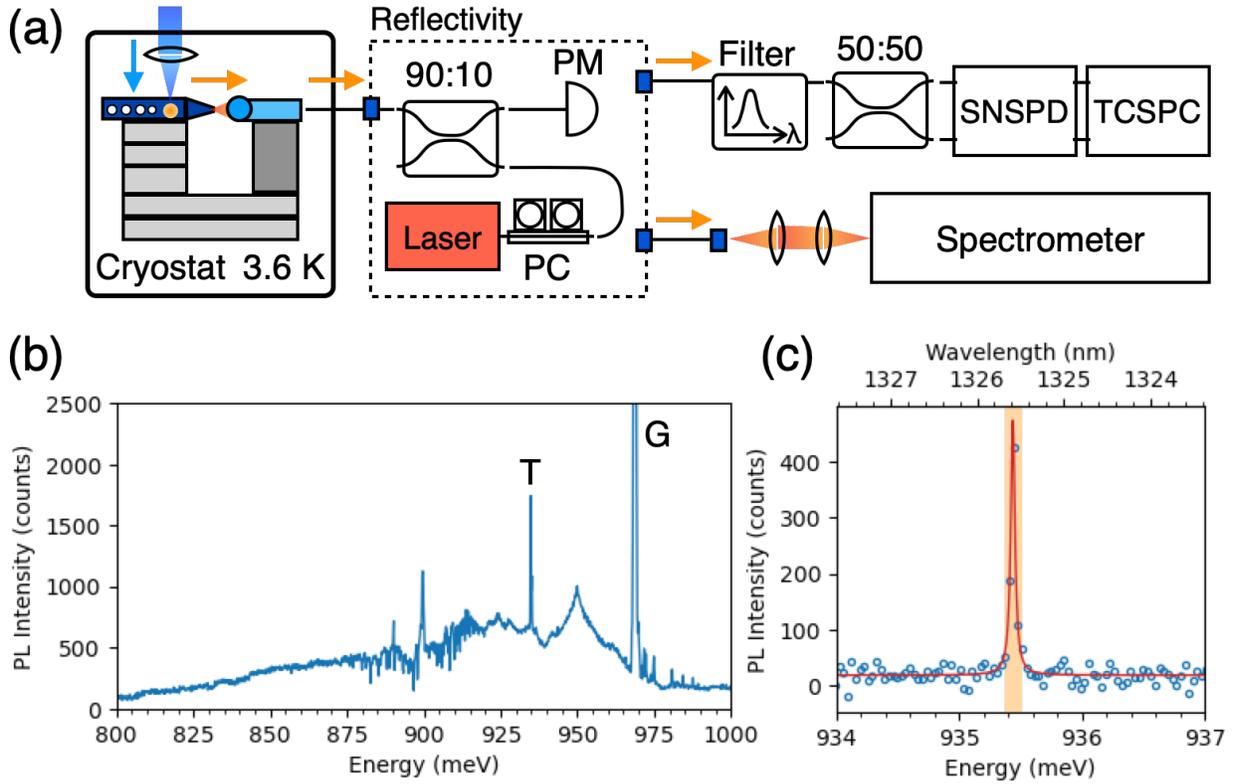

Figure 3 (a) A schematic of the optical measurement setup. Dashed box marked as 'Reflectivity' is used to measure reflectivity of the tapered nanobeam through lensed fiber. PM: power meter, PC: polarization controller, SNSPD: superconducting nanowire single photon detector, TCSPC: time-correlated single photon counter. (b) Above-band photoluminescence spectrum obtained by exciting the entire nanobeam with high pump power of 22 µW. The peaks at 935 meV and 968 meV correspond to the T center and the G center, respectively. (c) Above-band photoluminescence with tightly focused excitation beam spot at pump power of 1 µW. The red line corresponds to a Lorentzian fit of the measured data. The orange rectangle depicts a 0.2-nm spectral window that we used for single photon counting measurement.

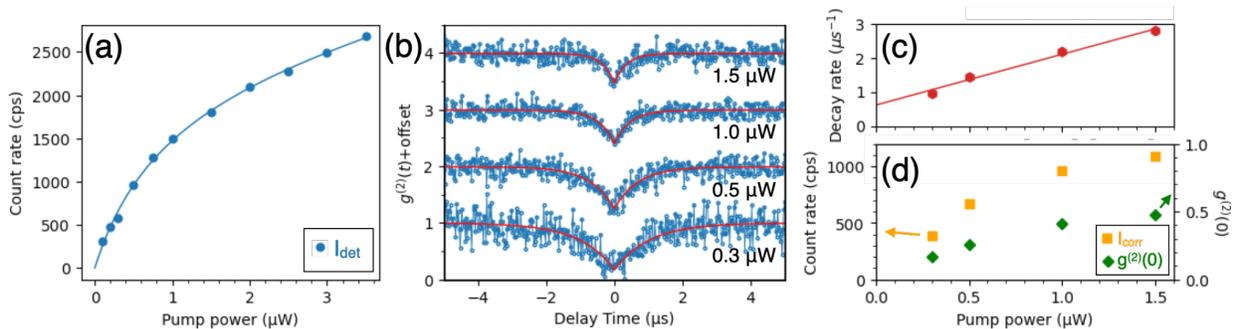

Figure 4 (a) Detected count rates as a function of pump power using the filtered zero phonon line of the T center. Solid line corresponds to a fit result modeled with a characteristic two level emitter saturation. cps: counts per second. (b) Second-order autocorrelation histograms at different pump powers. Red lines are the curves fitted to the double-sided exponential decay function. (c) Decay rate constant $\gamma_c$ at different pump powers obtained from the $g^{(2)}(\tau)$ histogram fits in panel (b). Solid line is a fit to a linear function. (d) Yellow squares are photon count rates corrected with the $g^{(2)}(0)$ values, green diamonds are the measured $g^{(2)}(0)$ values.